\documentclass[5p,number,sort&compress]{elsarticle}
\usepackage{amsmath,amssymb}
\usepackage{graphicx}
\usepackage[breaklinks,colorlinks=true]{hyperref}
\usepackage{slashed}

\newcommand{\MeV}{~\text{MeV}}
\newcommand{\GeV}{~\text{GeV}}
\newcommand{\Tc}{T_{\rm c}}

\newcommand{\LQCD}{\Lambda_{\text{QCD}}}

\newcommand{\Nc}{N_{\text{c}}}
\newcommand{\Nf}{N_{\text{f}}}

\newcommand{\bJ}{\boldsymbol{J}}
\newcommand{\bL}{\boldsymbol{L}}
\newcommand{\bS}{\boldsymbol{S}}

\newcommand{\boldomega}{\boldsymbol{\omega}}

\begin{document}

\title{
  \vspace*{-2em} {\small \hfill KEK-TH-2290, J-PARC-TH-0236} \\
  \vspace{1.5em}
  Deconfining Phase Boundary of Rapidly Rotating Hot and Dense Matter
  and Analysis of Moment of Inertia}

\author[UT]{Yuki~Fujimoto}
\ead{fujimoto@nt.phys.s.u-tokyo.ac.jp}
\author[UT]{Kenji~Fukushima}
\ead{fuku@nt.phys.s.u-tokyo.ac.jp}
\author[KEK,Sokendai,Riken]{Yoshimasa~Hidaka}
\ead{hidaka@post.kek.jp}

\address[UT]{Department of Physics, The University of Tokyo, %
  7-3-1 Hongo, Bunkyo-ku, Tokyo 113-0033, Japan}
\address[KEK]{Institute of Particle and Nuclear Studies, KEK, %
1-1 Oho, Tsukuba, Ibaraki 305-0801 Japan}
\address[Sokendai]{Graduate University for Advanced Studies (Sokendai), Tsukuba 305-0801, Japan}
\address[Riken]{RIKEN iTHEMS, RIKEN, 2-1 Hirosawa, Wako, Saitama 351-0198, Japan}

\begin{abstract}
  We discuss the effect of rapid rotation on the phase diagram of
  hadronic matter.  The energy dispersion relation is shifted by an
  effective chemical potential induced by rotation.  This suggests
  that rotation should lower the critical temperature of chiral
  restoration, but it is still controversial how the deconfinement
  temperature should change as a function of angular velocity.  We
  adopt the hadron resonance gas model as an approach free from
  fitting parameters.  We identify the deconfinement from the
  thermodynamic behavior and find that rotation decreases the
  deconfinement temperature.  We also discuss the spatial
  inhomogeneity of the pressure and give a semi-quantitative estimate
  of the moment of inertia.
\end{abstract}
\begin{keyword}
  hadronic matter \sep quarks and gluons \sep
  phase transition \sep deconfinement \sep rotation
\end{keyword}
\maketitle

\section{Introduction}

Rotation effects are ubiquitous in various systems and have been one
of the central topics in nuclear physics.  A prominent example is the
successful classification of nuclear spectra by the collective
rotational modes.  Some heavy nuclei spontaneously break rotational
symmetry by deformation and rotate to restore the broken symmetry
leading to the rotational band.  A classic review is found in
Ref.~\cite{Bohr:1976zz};  see also Ref.~\cite{Otsuka:2019diq} for a
modernized picture.

The present work addresses a nuclear system with more rapid rotation
and higher temperature created in non-central relativistic heavy-ion
collisions (see
Refs.~\cite{Fukushima:2018grm, Becattini:2020ngo, Huang:2020dtn}
for recent reviews).  In the experiment a nonzero value of $\Lambda$
and $\bar\Lambda$ polarization has been confirmed and the observed
polarization is translated to an angular velocity or vorticity
$\omega$ of created matter as large as
$\omega \simeq (9\pm1) \times 10^{21}~\text{s}^{-1}$~\cite{STAR:2017ckg}.
Some theoretical studies imply even higher values for
vorticity~\cite{Becattini:2015ska, Jiang:2016woz,Deng:2016gyh}.
Although expected $\omega$ values are extraordinarily large, the
corresponding energy scale is $\omega\sim 6\MeV$ in natural units.
Still, in addition to the temperature $T$, the baryon chemical
potential $\mu$, and the magnetic field $B$, the vorticity $\omega$
plays a role as a relevant parameter to characterize the properties of
hot and dense matter in heavy-ion collision phenomenology.  Now, a lot
of theoretical efforts are aimed to establish a full dynamical
description of the polarization in terms of hydrodynamics or kinetic
theory (see Refs.~\cite{Becattini:2020ngo, Huang:2020dtn} and
references therein).

Apart from phenomenology, the rotation has been also interested from
the wider point of view of quantum field theory or quantum
chromodynamics (QCD).  A well-known example is an anomalous transport
phenomenon; namely, the chiral vortical
effect~\cite{Vilenkin:1978hb, Vilenkin:1979ui, Vilenkin:1980zv, Son:2009tf}.
Another interesting possibility is an exotic ground state such as the
charged pion condensation emerging from a combination of rotation and
magnetic field~\cite{Liu:2017spl, Liu:2017zhl}.  Then, it is a natural
anticipation that $\omega$ should be a useful probe to investigate the
QCD phase diagram.  In theory $\omega$ could be chosen to be
comparable to a typical QCD scale, and it would be a quite inspiring
question how the QCD phase diagram evolves with increasing $\omega$.
In fact, the chiral phase transition has been already examined
extensively in the
literature~\cite{Chen:2015hfc, Jiang:2016wvv, Ebihara:2016fwa,
  Chernodub:2016kxh,Chernodub:2017ref, Wang:2018sur, Zhang:2018ome}.
It is a more or less accepted consensus that the rotation effect
suppresses the chiral condensate just like the finite density effect,
so that the chiral critical temperature drops with increasing $\omega$.

While most of the works have been concentrated on the chiral aspects,
the deconfinement transition in QCD is recently being
focused~\cite{Braguta:2020eis, Chen:2020ath, Chernodub:2020qah}.
One of the latest lattice-QCD calculations, which builds upon the
formulation in Ref.~\cite{Yamamoto:2013zwa}, claims that the
deconfinement temperature, $\Tc$, increases with growing $\omega$ by
measuring the Polyakov loop on the lattice in rotating
frames~\cite{Braguta:2020eis}.  A holographic QCD approach, by
contrast, suggests the opposite behavior, i.e., $\Tc$ decreases with
growing $\omega$~\cite{Chen:2020ath}, which is in accordance with the
behavior of the chiral critical temperature.  There is also an
alternative proposal of a mixed inhomogeneous phase supporting
spatially separated confinement and deconfinement
sectors~\cite{Chernodub:2020qah}.

In the present work we perform a simple yet robust analysis based on
the hadron resonance gas (HRG) model to estimate thermodynamic
quantities in a rotating frames.  This model has no free parameter
adjustable by hand and the input variables are all fixed by
experimentally observed particle spectra.  The virtue of the HRG model
lies in its unambiguousness of the minimal model definition.  The HRG
model (or the thermal model fit) has manifested eminent successes in
reproducing  the particle abundances in heavy-ion collision
experiments~\cite{Andronic:2017pug}.  Furthermore, the HRG model has
been found to be consistent with thermodynamic properties measured
in lattice-QCD simulations up to $\sim \Tc$ or even for higher $T$
once interactions are
included~\cite{Andronic:2012ut, Vovchenko:2014pka}.
The ideal (i.e., non-interacting) HRG model prevails as long as we stay
below $\Tc$, but for $T>\Tc$ the thermodynamic quantities predicted
from the ideal HRG model blow up.  The breakdown of such a hadronic
model based on the Hagedorn picture~\cite{Hagedorn:1965st} should be
identified as the deconfinement transition~\cite{Cabibbo:1975ig}.  We
will discuss this characterization later (see
Ref.~\cite{Andronic:2009gj} for more details).

To this end we formulate how to calculate the pressure in the rotating
frame.  With global rotation the pressure is inhomogeneous to be
balanced with the centrifugal force, from which we can infer the
distribution of the angular momentum and also the moment of inertia of
hot and dense matter.  Our numerical results agree with empirical
dependence on the radial distance from the rotation axis, and we make
a consistency check in a semi-quantitative way.

This paper is organized as follows.
In Sec.~\ref{sec:causality} we will briefly review the field
theoretical treatment of rotation and the energy dispersion relation
gapped by the causality bound.
In Sec.~\ref{sec:deconf} we will set forth our strategy to describe
deconfinement within the HRG model based on the Hagedorn picture.
In Sec.~\ref{sec:hrg} we will give an explicit expression for the
pressure in the rotating frame and spell out calculational procedures.
The pressure has explicit dependence on the radial distance from the
rotation axis and we closely discuss the physical interpretation in
Sec.~\ref{sec:radial}.
Section~\ref{sec:souzu} constitutes our central results in this paper
and we will show a 3D phase boundary surface as a function of the
baryon chemical potential $\mu$ and the angular velocity $\omega$.
In Sec.~\ref{sec:revisit} we will revisit the $r$ dependence to make a
consistency check between the physical interpretation and the
numerical results.
Section~\ref{sec:matome} is devoted to the summary and outlooks.

\section{Causality bound in the rotating frame}
\label{sec:causality}

The most straightforward approach to treat rotating systems is to
describe physics in a rotating frame by transforming non-rotating
coordinates, $\bar{x}^\mu$, into $x^\mu$ rotating with the angular
velocity $\omega$.  We take the rotation axis along the $z$ direction,
so that local quantities in the rotating frame are given as functions
of
\begin{equation}
  x = \bar{x}\cos\omega t + \bar{y}\sin\omega t\,,\qquad 
  y = -\bar{x}\sin\omega t + \bar{y}\cos\omega t\,.
  \label{eq:rot}
\end{equation}
We can read the metric as
\begin{equation}
  g_{\mu\nu}=\eta_{ab}
  \frac{\partial \bar{x}^a}{\partial x^\mu}
  \frac{\partial \bar{x}^b}{\partial x^\nu}
  = \begin{pmatrix}
    1-(x^2+y^2)\omega^{2} & y\omega & -x\omega & 0 \\
    y\omega & -1 & 0 & 0 \\
    -x\omega & 0 & -1 & 0 \\
    0 & 0 & 0 & -1
  \end{pmatrix}\,.
\end{equation}
Here, $\eta_{ab}$ represents the Minkowskian metric:\\
$\eta=\mathrm{diag}(1,-1,-1,-1)$.  For the fermionic particles the
equations of motion involve the first order derivatives, so we need
further to introduce the vierbein,
$\eta_{ab}= e^\mu_{\;a} e^\nu_{\;b} g_{\mu\nu}$, where
\begin{equation}
  e^t_{\;0} = e^x_{\;1} = e^y_{\;2} = e^z_{\;3} = 1\,\quad
  e^x_{\;0} = y\omega,\quad
  e^y_{\;0} = -x\omega\,,
\end{equation}
and the other components are zero.  The explicit
calculations~\footnote{We can find explicit calculations for
  $S=0$ and $S=1/2$ in the literature:
  see Ref.~\cite{Vilenkin:1980zv,Ambrus:2014uqa,Chen:2015hfc,Ebihara:2016fwa,Jiang:2016wvv}
  for several examples, and in the present work we assume that the
  energy shift by $-\bJ\cdot\boldomega$ generally holds for $S=1, 3/2$
  and $2$ as well.}
using the metric and the vierbein lead to a shift in the Hamiltonian as
\begin{equation}
  \hat{H} \;\to\; \hat{H} - \bJ\cdot\boldomega
  \label{eq:Hshift}
\end{equation}
in the rotating frame, where $\bJ$ is the total angular momentum;
namely, $\bJ=\bL+\bS$, with the orbital part $\bL$ and the spin part
$\bS$.  Accordingly the energy dispersion relation for spin-$S$
particles should be shifted as
\begin{equation}
  \varepsilon \;\to\; \varepsilon - (\ell + s)\omega\,.
\end{equation}
Here, $s=-S, -S+1,\dots S-1, S$ and $\ell$ denotes the quantum number
corresponding to the $z$ component of the orbital angular momentum,
i.e., $L_z$.

The energy shift above is analogous to the chemical potential for
finite density systems:  $(\ell+s)\omega$ can be regarded as an
effective chemical potential.  Then, one might think that, for
$(\ell+s)\omega > \varepsilon$, a Bose-Einstein condensate should form
for bosons, or, a Fermi surface should appear for fermions.  This is,
however, unphysical because Eq.~\eqref{eq:rot} is merely a coordinate
transformation and the \text{vacuum} physics must not change, while
the rotation together with \text{external} effects such as the
temperature, the electromagnetic fields, etc could make physical
differences.

In fact, $(\ell+s)\omega > \varepsilon$ is prohibited by the causality
condition.  In the cylindrical coordinates $(r,\varphi,z)$, we can
perform the Bessel-Fourier expansion to define the modes with
corresponding momenta, $(k_r,\ell,k_z)$, where $\ell$ is nothing but
the orbital angular momentum.  We can then impose a boundary condition
at $r=R$, such that the wave-functions are normalized within $r\le R$.
The causality requires,
\begin{equation}
  R\,\omega \le 1\,.
  \label{eq:causality}
\end{equation}
We note that the right hand side in the above condition is the speed
of light, $c=1$, in the natural unit.  The boundary condition at $r=R$
makes the momenta discretized as
$k_r = \xi_{\ell, n}/R$ with $\xi_{\ell, n}$ being the $n$th zero of
the Bessel function:
$J_\ell(\xi_{\ell,n})=0$~\cite{Vilenkin:1980zv,Ambrus:2015lfr,Ebihara:2016fwa}.
It is important to note that $J_{\ell\ge 1}(\xi)$ has a zero at
$\xi=0$ but such a zero mode is identically vanishing due to the
boundary condition at $r=R$, and there is no zero mode contribution to
physical quantities.  Strictly speaking, thus, $\xi_{\ell,n}$ indicates
the $n$th zero excluding a trivial zero at $\xi=0$.  In this way we
can conclude that the energy is always gapped at least by
$\xi_{\ell,1}/R$ for any particles.  We combine this gap and
the condition~\eqref{eq:causality} to confirm,
\begin{equation}
  \varepsilon \;\ge\; \frac{\xi_{\ell,1}}{R} \;\ge\; \xi_{\ell,1}\,\omega\,.
\end{equation}
It is known that $\xi_{\ell,n}$'s satisfy the following inequality:
$\xi_{\ell,1} > \ell + 1.855757\ell^{1/3} + 0.5\ell^{-1/3}$
for $\ell\ge 1$ and $\xi_{0,1}=2.40483$.  These relations guarantee
$\varepsilon > (\ell+s)\omega$ for sufficiently large $\ell$ even when
$s$ is large.  Our present calculations, as we explain in details
later, contain hadrons up to $S=2$ and
$\varepsilon > (\ell+s)\omega$ holds for $|s|\le S$ and all $\ell$.

\section{Deconfinement transition in the Hagedorn picture}
\label{sec:deconf}

The HRG model analysis is thoroughly hadronic, but we can still
discuss the deconfinement transition as follows.  Historically
speaking, the Hagedorn limiting temperature was first recognized
within the framework of hadronic bootstrap
model~\cite{Hagedorn:1965st}.  Later, then, Cabibbo and Parisi
realized that the limiting temperature should be given a correct
physical interpretation as the transition temperature to more
fundamental degrees of freedom than hadrons~\cite{Cabibbo:1975ig}. 

Let us suppose that the hadron mass spectrum rises exponentially,
i.e.,
\begin{equation}
  \rho(m) = e^{m/T_H}\,,
  \label{eq:Hagedorn}
\end{equation}
where $T_H$ is not a physical temperature but just a slope parameter
to characterize the mass spectrum.  Then, the integration weighted
with the Boltzmann factor, $e^{-m/T}$, gives us the partition function
as
\begin{equation}
  Z = \int dm\, \rho(m)\, e^{-m/T}\,.
\end{equation}
For simplicity we omit the phase space volume (that would give a
polynomial factor) and focus on the exponential behavior only.  In
other words the integration measure of $dm$ is implicitly defined in a
consistent way.  Now, it is obvious that the integration diverges for
$T>T_H$, and Hagedorn considered that $T_H$ should be the limiting
temperature:  any physical systems of hadrons cannot be heated above
$T_H$.  This conjecture should be revised once internal structures of
hadrons are taken into account.  The existence of $T_H$ should be
correctly interpreted as a breakdown point of such a simple hadronic
description and the physical systems should be better characterized by
quarks and gluons at $T>T_H$.

In the HRG model, the hadron mass spectrum is taken from the
experimental data, and interestingly, $\rho(m)$ shows exponential
growth up to $m\sim 3\GeV$.  Therefore, the above picture of
deconfinement makes approximate sense, and we can see blowup behavior
of thermodynamic quantities such as the pressure, the internal energy,
the entropy density, and so on at a certain temperature ($T\sim T_H$),
though they do not diverge strictly.  Therefore, we can physically
identify the deconfinement crossover point from the blowup behavior of
thermodynamic quantities in the HRG model.  It is straightforward to
extend the above mentioned picture to a finite density case by
replacing the Boltzmann factor with $e^{-(m-\mu)/T}$ for baryons that
also exhibit exponential spectra $\sim e^{m/T_B}$ (see
Ref.~\cite{Andronic:2009gj}).  We will explain our working criterion
for deconfinement in later discussions.

\section{Rotating hadron resonance gas model}
\label{sec:hrg}

The HRG model has been well established and for our purpose to
investigate rotating systems we need to rewrite the formulas in terms
of the cylindrical coordinates, $(k_r, \ell, k_z)$.  The pressure in
the HRG model has contributions from both mesons ($m$) and baryons
($b$) up to an ultraviolet mass scale, $\Lambda$:
\begin{equation}
  p(T, \mu, \omega; \Lambda) = \sum_{m;\,M_i\le \Lambda} p_m
  + \sum_{b;\,M_b\le \Lambda} p_b\,,
\end{equation}
The mesonic and the baryonic pressures are given by
\begin{equation}
  p_m = p_{i=m}^-\,,\qquad
  p_b = p_{i=b}^+\,,
\end{equation}
where the generalized pressure functions are
\begin{align}
  p_i^{\pm}
  &= \pm\frac{T}{8\pi^2} \sum_{\ell=-\infty}^\infty \int dk_r^2
    \int  dk_z\; \sum_{\nu = \ell}^{\ell + 2S_i}  J_\nu^2(k_r r) \notag\\
  &\qquad\qquad
    \times \log\left\{1\pm\exp[-(\varepsilon_{\ell,i} - \mu_i) / T]\right\}\,. 
  \label{eq:hrgrot}
\end{align}
The energy spectrum is
$\varepsilon_{\ell, i} = \sqrt{k_r^2 + k_z^2 + m_i^2} - (\ell + S_i) \omega$
with $S_i$ and $m_i$ being the spin and the mass of the particle $i$.
We note that the radial integration is with respect to $k_r^2$ in the
above form; that is, $dk_r^2=2k_r dk_r$.
The above expression needs some more explanations.  The rotation
effect shifts the energy dispersion relation by the cranking term,
i.e., $ - \bJ \cdot \boldomega$, which varies as $(\ell+s_i)\omega$
from $s_i=-S_i$ to $s_i=+S_i$.  We reorganize the sum over $s_i$ and
$\ell$ so that the energy shift can be the same, $-(\ell+S_i)\omega$,
to simplify the expression.  Then, the spin sum is translated to the
sum with respect to $\nu$ with the square of the Bessel function $J_\nu^2(k_r r)$ as
in Eq.~\eqref{eq:hrgrot}.  The Bessel function arises from the weight
in the Bessel-Fourier expansion.  The simplest nontrivial example is
the spin-1/2 calculation (see Ref.~\cite{Ebihara:2016fwa,Chen:2017xrj}
for more details).  After the appropriate redefinition of $\ell$ in
such a way that the total angular momentum is $j=\ell+1/2$, one
particle solutions of the Dirac equation read:
\begin{equation}
  u_+ = \frac{e^{-i\varepsilon t + ik_z z}}{\sqrt{\varepsilon + m}}
  \begin{pmatrix}
    (\varepsilon+m) J_\ell (k_r r) e^{i\ell\varphi} \\
    0 \\
    k_z J_\ell(k_r r) e^{i\ell\varphi} \\
    ik_r J_{\ell+1}(k_r r) e^{i(\ell+1)\varphi}
  \end{pmatrix}\,.
\end{equation}
The other solution, $u_-$, can be expressed similarly (the explicit
expression is found in Ref.~\cite{Ebihara:2016fwa}).  From these
solutions the fermionic propagator can be constructed and its trace
involves $J_\ell^2(k_r r) + J_{\ell+1}^2(k_r r)$, that is nothing but
the sum we see in Eq.~\eqref{eq:hrgrot} for $S_i=1/2$.

It is important to note that the integrations and the sum in
Eq.~\eqref{eq:hrgrot} are convergent.  We can understand that from the
$\omega\to0$ limit to recover the standard expression in the HRG model:
\begin{equation}
  p_i^{\pm} \to \pm\frac{g_iT}{2\pi^2}\int_0^\infty \!k^2dk 
  \log\left\{1\pm\exp\left[-\frac{\sqrt{k^2 \!+\! m_i^2}
        \!-\! \mu_i}{T}\right]\right\}\,,
  \label{eq:ihrg}
\end{equation}
where $g_i=2S_i+1$ is the spin degeneracy factor and this expression is
certainly convergent.  The dispersion relation involves an
exponentially growing factor, $e^{\ell\omega/T}$, but
$J_{\nu\ge \ell}^2(k_r r)$ has stronger exponential suppression and
Eq.~\eqref{eq:hrgrot} is finite.

There is, however, one subtlety in Eq.~\eqref{eq:hrgrot}.  As
discussed in Sec.~\ref{sec:causality}, we can avoid unphysical
condensates from the causality bound, but it is time consuming to take
the discrete sum of $k_r$.  Here, instead, we shall employ an
approximate and minimal prescription to evade unphysical condensates.
As long as $\omega$ is not significantly larger than $\LQCD$, the
discretization in high momentum regions is expected to be a minor
effect, and the leading discretization effect in the low momentum
regions is the mass gap.  We can thus introduce an infrared cutoff for
the $k_r$ integration, $\Lambda^{\text{IR}}_\ell$, defined by 
\begin{equation}
  \Lambda^{\text{IR}}_\ell = \xi_{\ell, 1} \omega\,, 
\end{equation}
where, as we already noted, an obvious zero at $\xi=0$ is excluded.
The $k_r$ integration in Eq.~\eqref{eq:hrgrot} is then replaced as
\begin{equation}
  \int dk_r^2 \quad\to\quad \int_{(\Lambda^{\text{IR}}_\ell)^2} dk_r^2\,.
\end{equation}
We will elucidate technical procedures in more details in
Sec.~\ref{sec:souzu}.

\section{Radial dependence}
\label{sec:radial}

We note that our main formula~\eqref{eq:hrgrot} depends on the radial
coordinate $r$ through $J_\nu^2(k_r r)$.  There are twofold intuitive
origins for this $r$ dependence.  One is possible $r$ dependence from
the boundary effect at $R\sim 1/\omega$.  The boundary effect exists
even for non-rotating matter.  We are interested in not surface
singularities (as discussed in Ref.~\cite{Chen:2017xrj} for example)
but bulk properties, and so we can take as small $r$ as possible for
numerical implementation.  Another origin is that the centrifugal
force should be supported by the $r$ dependent part of the pressure.

Let us consider the $r$ dependence from the latter origin.  From the
analogy to the relation between the baryon number density and the
pressure: $n=\partial p/\partial\mu$, we can express the angular
momentum density as
\begin{equation}
  \langle j\rangle(r) = \frac{\partial p(r)}{\partial \omega}\,.
\end{equation}
When $\omega$ is small in the linear regime, the angular momentum is
related to the moment of inertia in the infinitesimal volume $dV$ as
\begin{equation}
  \langle j\rangle(r) \, dV \simeq dI(r) \omega\,.
\end{equation}
For homogeneous matter with mass density $\rho$, we can easily find
the moment of inertia as $dI(r) = \rho r^2\, dV$. If the
baryon chemical potential is vanishing, $\rho$ should be characterized
by the temperature $T$, i.e., $\rho=\sigma T^4$.  We can roughly
approximate $\sigma$ from the enthalpy density; namely,
$\sigma=2\nu \pi^2/45$ with the thermal degrees of freedom $\nu$.
Then, we can approximate:
\begin{equation}
  p(r) = p(0) + \Delta p(r)\,,\qquad
  \Delta p(r) \simeq \frac{\sigma}{2} T^4 r^2 \omega^2\,.
\label{eq:rdep}
\end{equation}
Because $\sigma$ may differ for confined hadronic matter and
deconfined matter of quarks and gluons, the deconfinement point could
be in principle dependent on $r$.  Indeed in the cylinder with a
boundary, the possibility of spatially separated regions of
confinement and deconfinement was pointed
out~\cite{Chernodub:2020qah}.

In the present work, to avoid ambiguous interpretation, we shall take
$r \omega \ll 1$ so  that we can safely neglect the $r$ dependence:
we fix $r=0.01\GeV^{-1}$ throughout this work.  If we take the strict
limit of $r\to 0$ in the integrand in Eq.~\eqref{eq:hrgrot} (assuming
that the infinite sum over $\ell$ and the integration with respect to
$k_r$ are harmless), all the terms involving $J_{\nu\neq 0}^2(0)=0$
should vanish.  Then, only terms with $\nu=0$ survive, which are
allowed for $\ell=-2S_i$ to $\ell=0$, corresponding to the energy
shifts from $-S_i\omega$ to $+S_i\omega$.  Since we redefined $\ell$
to simplify Eq.~\eqref{eq:hrgrot}, it is a bit nontrivial to see, but
the surviving terms are different spin states with zero orbital
angular momentum.  This is very natural:  at $r=0$ the orbital angular
momentum is identically zero and the rotation couples to the spin
only.

\section{Numerical results}
\label{sec:souzu}

\begin{figure*}
  \includegraphics[width=.32\textwidth]{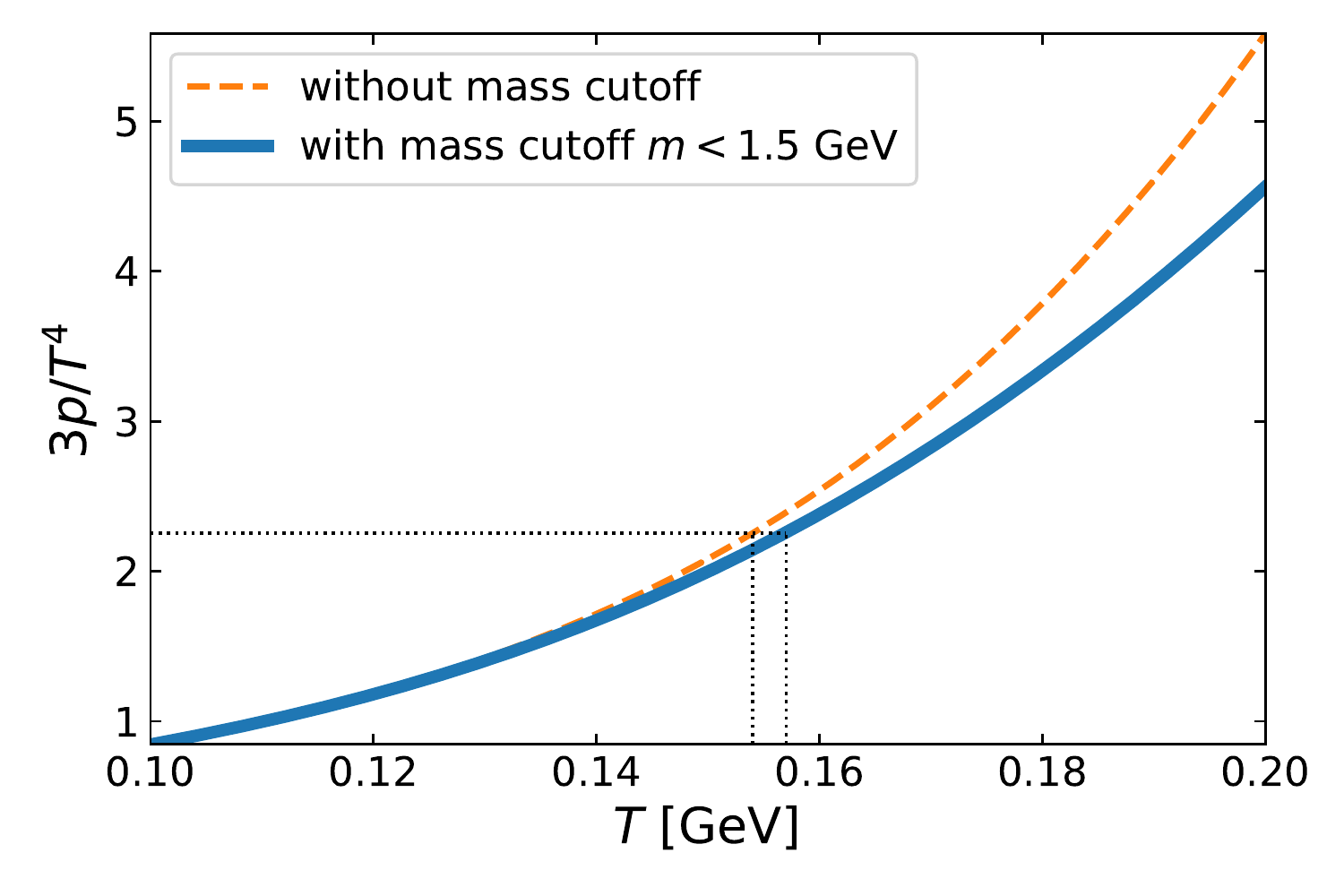}
  \includegraphics[width=.32\textwidth]{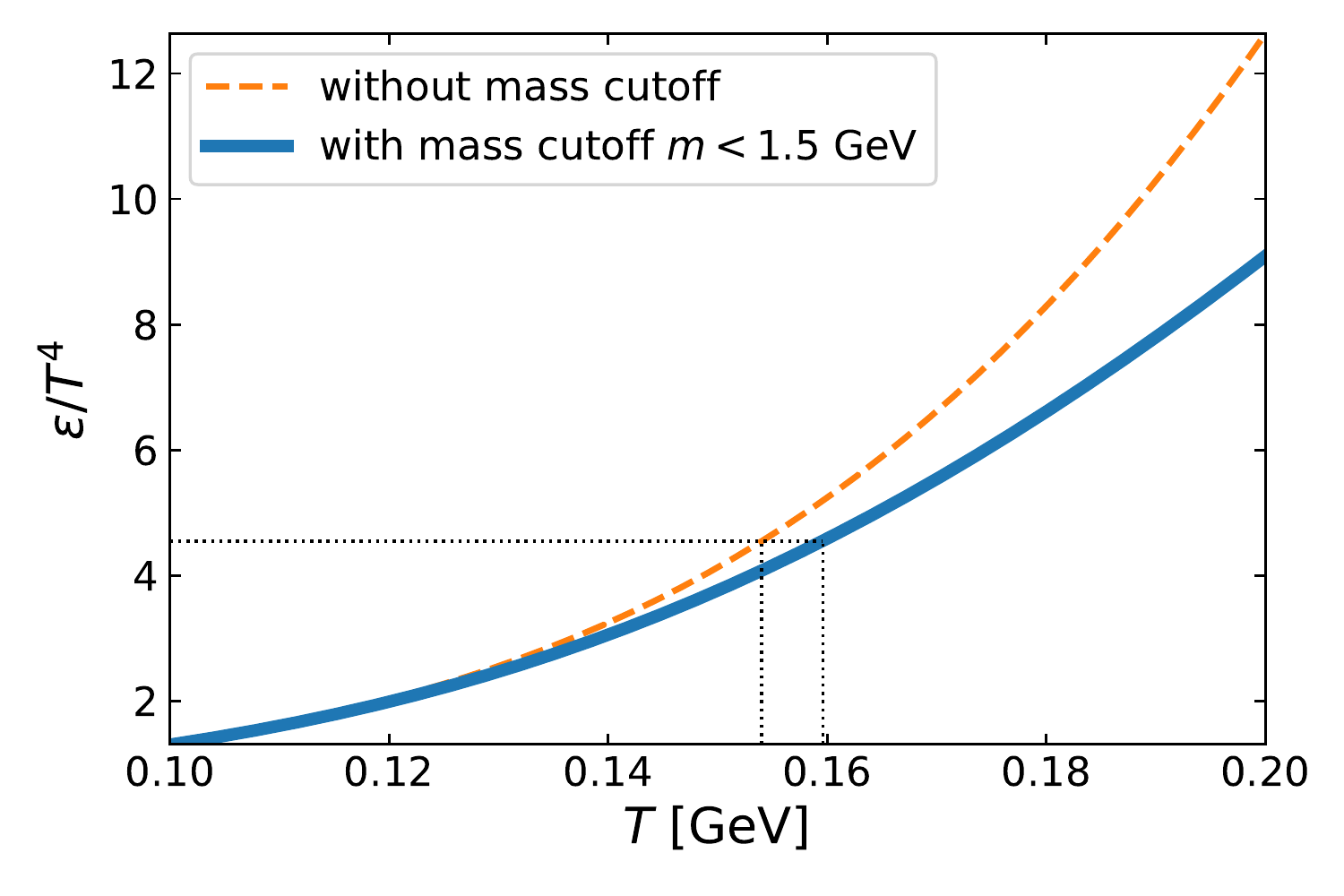}
  \includegraphics[width=.32\textwidth]{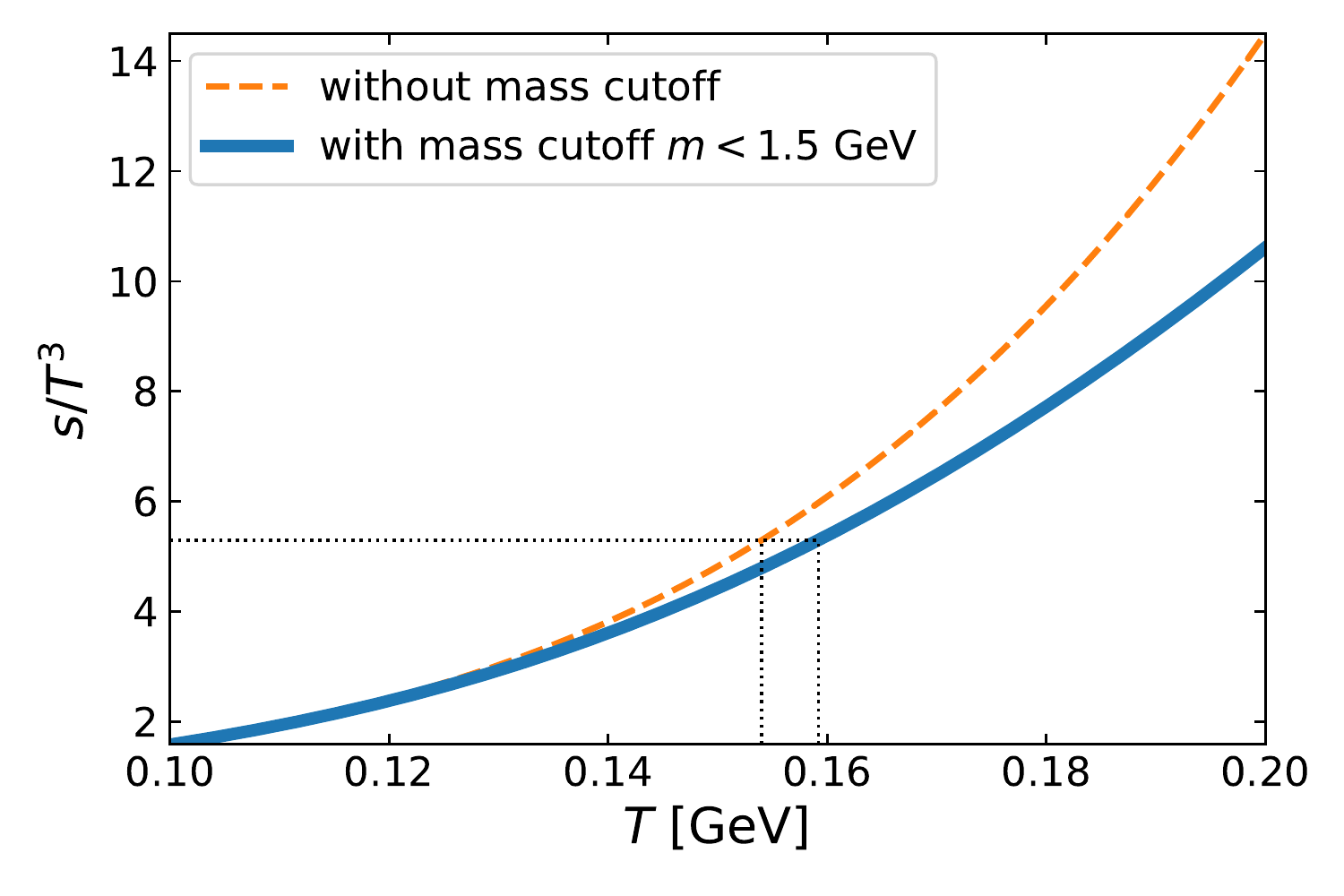}
  \caption{Thermodynamic quantities, the pressure (left), the energy 
    density (middle), and the entropy (right), calculated in the HRG 
    model with and without imposing the mass cutoff $m<\Lambda$ with 
    $\Lambda=1.5\GeV$.}
  \label{fig:ihrg}
\end{figure*}

In our HRG model treatment we have adopted the particle data group
list of particles contained in the package of
THERMUS-V3.0~\cite{Wheaton:2004qb} and incorporated the data into our
own numerical codes.  To reduce the numerical cost, we impose an
ultraviolet mass cutoff as $\Lambda=1.5\GeV$ in Eq.~\eqref{eq:hrgrot}.
This also limits the high spin states.  With our choice of
$\Lambda=1.5\GeV$ the largest spin states contributing to the pressure
are $f_2(1270)$, $a_2(1320)$, $K_2^\ast(1430)$, and $f_2(1430)$
with $S=2$.  The effect of $\Lambda$ on the chemical freezeout curve
has been examined in Ref.~\cite{Cleymans:2005xv}, and they have found
that the changes of the chemical freezeout curve are as small as
around $10\MeV$.

We quantitatively study the effect of $\Lambda$.  In
Fig.~\ref{fig:ihrg} we plot the thermodynamic quantities with and
without the cutoff from Eq.~\eqref{eq:ihrg} in the standard
non-rotating HRG model.  The left panel shows the pressure $p$, the
middle shows the energy density $\varepsilon$, and the right shows the
entropy density $s$ as functions of $T$.  To check the validity of our
simplification with $\Lambda$, we shall compare the critical
temperature $\Tc$ read out from a thermodynamic criterion.

The critical temperature without $\Lambda$ is known from the
lattice-QCD simulation as $\Tc = 154\MeV$~\cite{Bazavov:2014pvz}.  We
can find the corresponding critical $p/T^4$, $\varepsilon/T^4$, and
$s/T^3$ at $\Tc$ from the crossing points of the orange dashed curves
and the dotted vertical lines.  Then, we can estimate the $\Lambda$
modified $\Tc$ from the crossing points of the blue solid curves and
the dotted horizontal lines in Fig.~\ref{fig:ihrg}.  The shifts in
$\Tc$ read out from $p/T^4$, $\varepsilon/T^4$, and $s/T^3$ are
$3.0\MeV$, $5.6\MeV$, and $5.2\MeV$, respectively.  This is the
numerical confirmation that the $\Lambda$ effects on $\Tc$ are less
than $10\MeV$.  In conclusion, our simplification by $\Lambda=1.5\GeV$
is qualitatively harmless for the study of the phase boundary around
$\Tc$ and also at the quantitative level the possible error is
$\sim 5\MeV$.  We assume that the $\Lambda$ effects are negligible for
finite $\omega$ as well.

Now let us discuss the deconfinement phase boundaries at finite $\mu$
and $\omega$.  For this purpose we should make the thermodynamic
quantities not only with $T$ (as in Fig.~\ref{fig:ihrg}) but with some
proper combination of $T$, $\mu$, and $\omega$.  We employ the
normalization given by the Stefan-Boltzmann limit of a rotating
quark-gluon gas:
\begin{equation}
  p_{\text{SB}} \equiv  (\Nc^2 - 1) \,p_{\rm g} + \Nc \Nf \,
  (p_{\rm q} + p_{\rm \bar{q}})\,,
\end{equation}
where the number of colors and flavors are $\Nc=3$, $\Nf=2$, respectively.
The gluon pressure reads:
\begin{align}
  p_{\rm g}
  &= -\frac{T}{8\pi^2}\sum_{\ell=-\infty}^\infty
    \int_{\Lambda^{\text{IR}}_\ell} dk_r^2 \int dk_z
    \left[J_{\ell}^2(k_r r) + J_{\ell+2}^2(k_r r)\right]\notag\\
  & \quad\times \log\left\{1
    -\exp\left[-\frac{\sqrt{k_r^2+k_z^2} - (\ell +1) \omega}{T}\right]\right\}\,.
\end{align}
Here, we note that the possible angular momenta are only $j=\ell-1$ and
$j=\ell+1$ and there is no contribution from $s_z=0$ because gluons
are massless gauge bosons.  This is why
$J_{\ell}^2(k_r r) + J_{\ell+2}^2(k_r r)$ appears above.  The quark
pressure reads more straightforwardly:
\begin{align}
  p_{\rm q}
  &= -\frac{T}{8\pi^2}\sum_{\ell=-\infty}^\infty
    \int_{\Lambda^{\text{IR}}_\ell} dk_r^2 \int dk_z
    \left[J_{\ell}^2(k_r r) + J_{\ell+1}^2(k_r r)\right]\notag\\
  & \quad\times
    \log\left\{1\!+\!\exp\left[-\frac{\sqrt{k_r^2\!+\!k_z^2}
    \!+\! (\ell \!+\!\frac{1}{2}) \omega \!-\! \frac{\mu}{\Nc}}{T}\right]\right\}
\end{align}
and the anti-quark pressure, $p_{\rm \bar{q}}$, takes almost the same
expression with $\mu\to -\mu$.

\begin{figure}
  \includegraphics[width=\columnwidth]{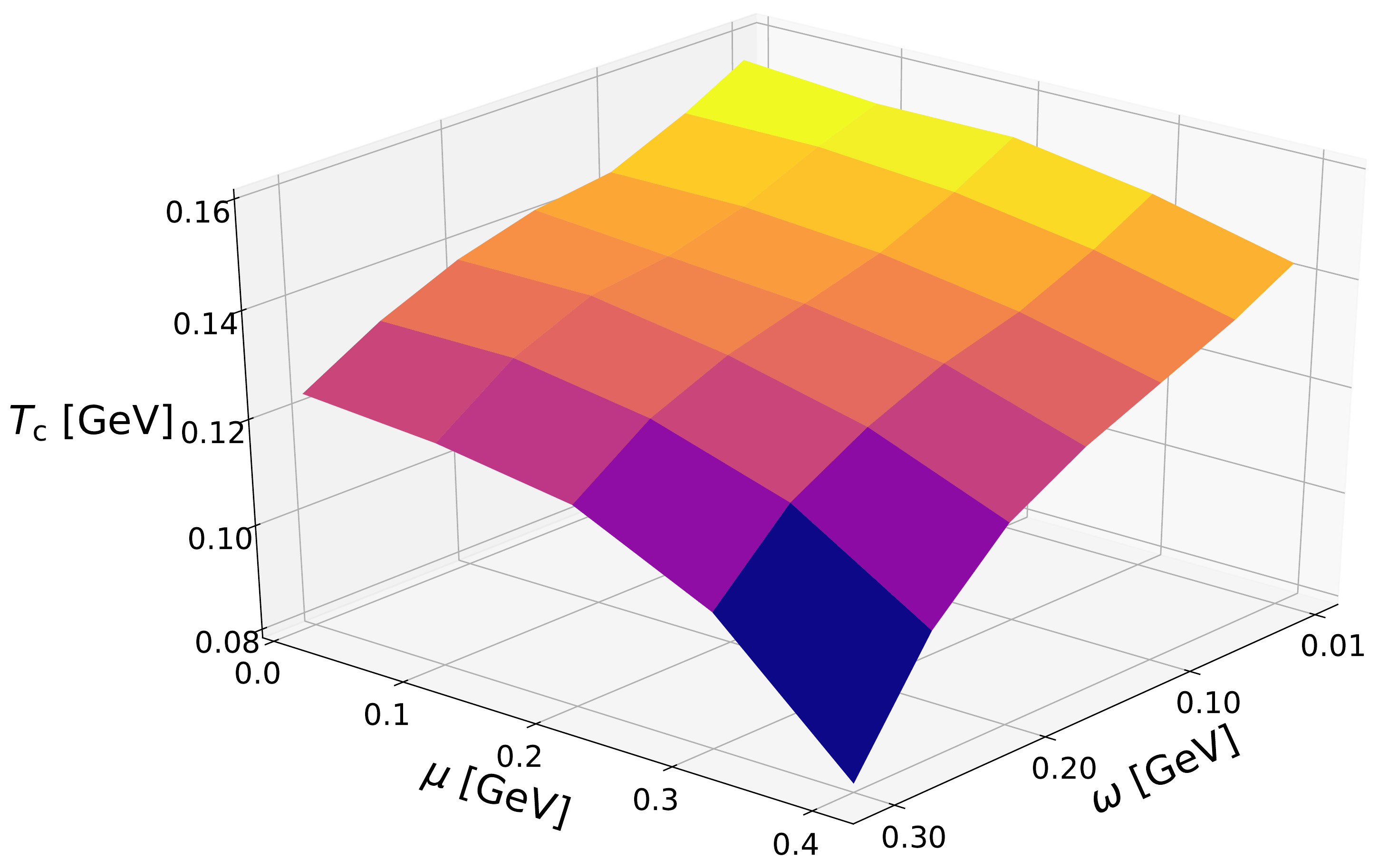}
  \caption{Deconfinement transition surface as a function of the 
    baryon chemical potential $\mu$ and the angular velocity 
    $\omega$.}
  \label{fig:3d}
\end{figure}

Here our criterion for the deconfinement transition is prescribed, in a
way similar to Ref.~\cite{Fukushima:2010is}, as
\begin{equation}
  \frac{p}{p_{\text{SB}}}(\Tc,\,\mu,\,\omega) = \gamma\,.
  \label{eq:criterion}
\end{equation}
Here, $\gamma$ is a constant, which is chosen to reproduce
$\Tc(\mu=\omega=0)=154\MeV$ in accordance with the lattice-QCD
results~\cite{Bazavov:2014pvz}.  This condition fixes $\gamma=0.18$ in
our calculation.  Now we can numerically solve
Eq.~\eqref{eq:criterion} to identify $\Tc=\Tc(\mu,\,\omega)$ as
plotted in Fig.~\ref{fig:3d}.

Now it is evident that $\Tc$ is a decreasing function with increasing
$\omega$ just like the behavior along the $\mu$ direction.  We cannot
directly study the chiral properties within the HRG model, but it is
conceivable that the deconfinement $\Tc$ and the chiral restoration
temperature are linked even at finite $\omega$.  We can also notice
that the effect of $\omega$ makes $\Tc$ drop faster than that of
$\mu$.  We understand this from the $\omega$ induced effective
chemical potential which is proportional to $\ell+S_i$.  Because
$\ell$ becomes arbitrarily large, the system can be more sensitive to
the effective chemical potential than the baryon chemical potential.
From our parameter free analyses we make a conclusion that the
deconfining transition temperature is lowered by the rotation effect.

\section{Revisiting the radial dependence}
\label{sec:revisit}

It would be an interesting problem to make systematic investigations
of the $r$ and $\omega$ dependence in the pressure.  The main focus of
the present work is the survey of the phase diagram, so we will not go
into systematic discussions here.  Still, it would be instructive to
verify our physical interpretation of the $r$ and $\omega$ dependence
in Eq.~\eqref{eq:rdep} from the numerical calculation.

\begin{figure}
  \includegraphics[width=\columnwidth]{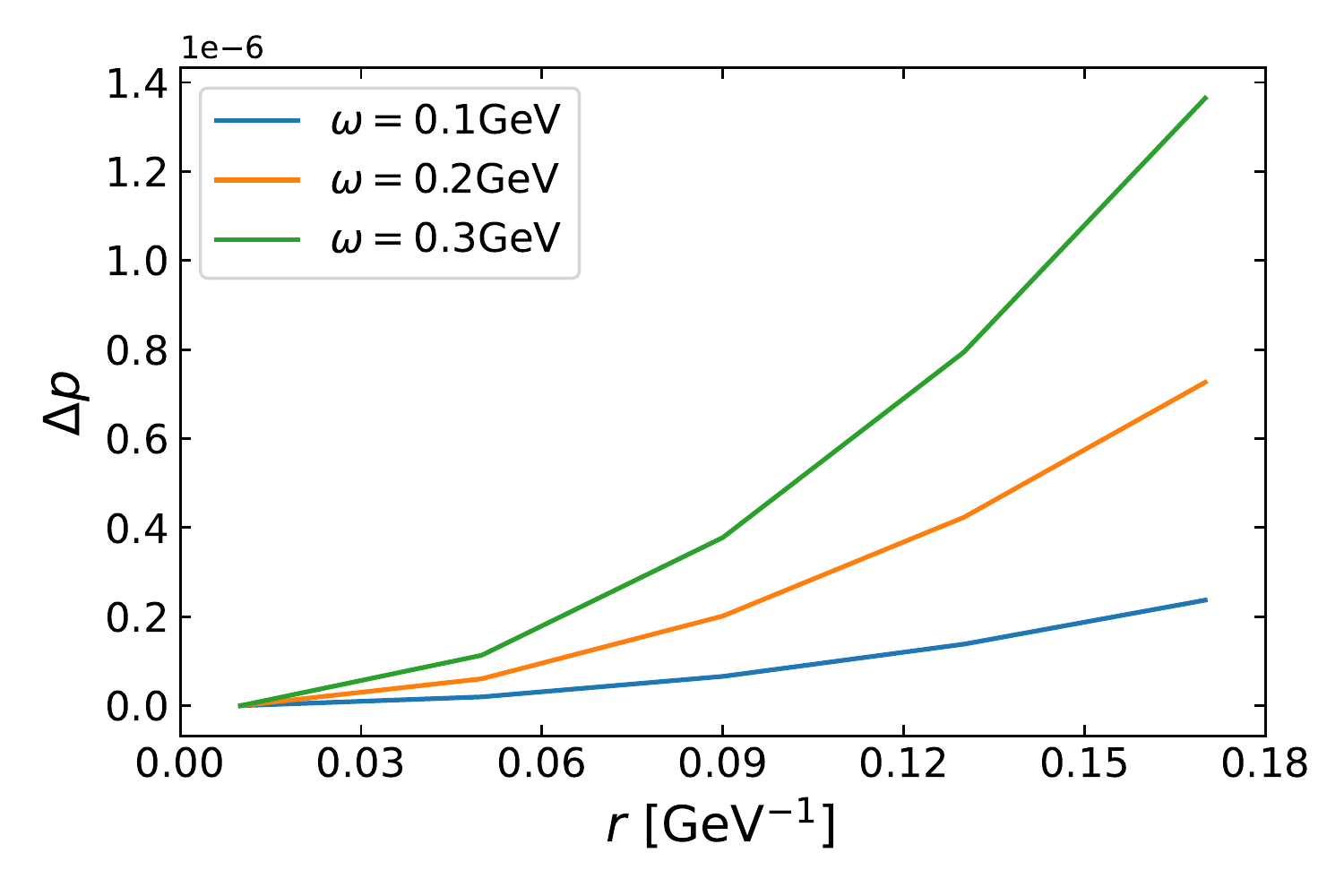}
  \caption{$\Delta p$ as a function of $r$ for three different values
    of $\omega$.}
  \label{fig:rdep}
\end{figure}

We fix the temperature, $T=0.15\GeV$, and change $r$ for three
different values of $\omega=0.1$, $0.2$, $0.3\GeV$.  The range of $r$
is $[0.01, 0.17]\GeV^{-1}$.  Our numerical calculations lead to the
$r$ dependence as shown in Fig.~\ref{fig:rdep}.  We have checked that
each curve on Fig.~\ref{fig:rdep} is well fitted by a quadratic
function $\propto r^2$ as expected from Eq.~\eqref{eq:rdep}.  From
this quadratic $r$ dependence we can numerically estimate $\sigma$
defined in Eq.~\eqref{eq:rdep}.  For $\omega=0.1\GeV$ the numerical
coefficient reads: $\Delta p/r^2 \simeq 8.19141\times 10^{-6}\GeV^{6}$.
The corresponding value of $\sigma$ is $\sigma\simeq 3.21$, from which
we can infer,
\begin{equation}
  \nu(\omega=0.1\GeV) \simeq 7\,.
\end{equation}
For different $\omega$ the results are slightly changed, but of the
same order.  This value of $\nu$ is comparable to the thermal degrees
of freedom of light mesons, i.e., pions and Kaons.  
We have a full expression of Eq.~\eqref{eq:hrgrot} and we do
not have to rely on an Ansatz like Eq.~\eqref{eq:rdep}.  In this sense
the above mentioned estimate of $\nu$ should be understood as a
consistency check.  It would be a very intriguing question to see the
spatial distribution of the angular momentum density,
$\langle j\rangle(r)$, as well as the moment of inertia, $dI(r)$,
directly from Eq.~\eqref{eq:hrgrot}.  We will report a thorough
analysis in a separate publication and stop our discussions at the
level of the consistency check in the present paper.

\section{Summary}
\label{sec:matome}

We studied the effect of rotation on the deconfinement transition from
hadronic to quark matter.  We devised the hadron resonance gas (HRG)
model in a rotating frame and formulated a practical scheme for the
pressure calculation that is dependent on the radial distance $r$ from
the rotation axis.  Adopting a working criterion for deconfinement in
the view of the Hagedorn picture, we found that increasing the angular
velocity $\omega$ lowers the deconfinement transition temperature,
which is similar to the effect of baryon chemical potential.  We then
drew the 3D phase diagram of rotating hot and dense matter in
Fig.~\ref{fig:3d}.  Our physics discussions include not only the phase
diagram but also the physical interpretation of the spatial dependence
of the pressure.  The numerical results are consistent with the
physical interpretation in terms of the moment of inertia.

There are many interesting directions for future extensions.  In the
context of the QCD phase diagram research it would provide us with an
inspiring insight to study whether the deconfinement and the chiral
restoration transitions should be locked or unlocked by rotational
effects.  Also, a more comprehensive analysis involving the magnetic
field on top of rotation would be desirable for phenomenological
applications.  Since we formulated the pressure as a function of $r$
and $\omega$, it have paved a clear path for the microscopic
computation of the angular momentum and the moment of inertia of
rotating hot and dense matter.  Such quantities should be valuable for
phenomenological modeling.  We are making progress along these lines
using perturbative QCD as well as the HRG model.

\section*{Acknowledgments}
This work was supported by Japan Society for the Promotion of Science
(JSPS) KAKENHI Grant
Nos.\ 18H01211 (KF,YH), 19K21874 (KF), 17H06462 (YH), and 20J10506 (YF).

\bibliography{bib_HRGrot}

\begin{thebibliography}{10}
\expandafter\ifx\csname url\endcsname\relax
  \def\url#1{\texttt{#1}}\fi
\expandafter\ifx\csname urlprefix\endcsname\relax\def\urlprefix{URL }\fi
\expandafter\ifx\csname href\endcsname\relax
  \def\href#1#2{#2} \def\path#1{#1}\fi

\bibitem{Bohr:1976zz}
A.~Bohr, {Rotational motion in nuclei}, Rev. Mod. Phys. 48 (1976) 365--374.
\newblock \href {https://doi.org/10.1103/RevModPhys.48.365}
  {\path{doi:10.1103/RevModPhys.48.365}}.

\bibitem{Otsuka:2019diq}
T.~Otsuka, Y.~Tsunoda, T.~Abe, N.~Shimizu, P.~Van~Duppen, {Underlying structure
  of collective bands and self-organization in quantum systems}, Phys. Rev.
  Lett. 123~(22) (2019) 222502.
\newblock \href {http://arxiv.org/abs/1907.10759} {\path{arXiv:1907.10759}},
  \href {https://doi.org/10.1103/PhysRevLett.123.222502}
  {\path{doi:10.1103/PhysRevLett.123.222502}}.

\bibitem{Fukushima:2018grm}
K.~Fukushima, {Extreme matter in electromagnetic fields and rotation}, Prog.
  Part. Nucl. Phys. 107 (2019) 167--199.
\newblock \href {http://arxiv.org/abs/1812.08886} {\path{arXiv:1812.08886}},
  \href {https://doi.org/10.1016/j.ppnp.2019.04.001}
  {\path{doi:10.1016/j.ppnp.2019.04.001}}.

\bibitem{Becattini:2020ngo}
F.~Becattini, M.~A. Lisa, {Polarization and Vorticity in the Quark Gluon
  Plasma} (3 2020).
\newblock \href {http://arxiv.org/abs/2003.03640} {\path{arXiv:2003.03640}},
  \href {https://doi.org/10.1146/annurev-nucl-021920-095245}
  {\path{doi:10.1146/annurev-nucl-021920-095245}}.

\bibitem{Huang:2020dtn}
X.-G. Huang, J.~Liao, Q.~Wang, X.-L. Xia, {Vorticity and Spin Polarization in
  Heavy Ion Collisions: Transport Models} (10 2020).
\newblock \href {http://arxiv.org/abs/2010.08937} {\path{arXiv:2010.08937}}.

\bibitem{STAR:2017ckg}
L.~Adamczyk, et~al., {Global $\Lambda$ hyperon polarization in nuclear
  collisions: evidence for the most vortical fluid}, Nature 548 (2017) 62--65.
\newblock \href {http://arxiv.org/abs/1701.06657} {\path{arXiv:1701.06657}},
  \href {https://doi.org/10.1038/nature23004} {\path{doi:10.1038/nature23004}}.

\bibitem{Becattini:2015ska}
F.~Becattini, G.~Inghirami, V.~Rolando, A.~Beraudo, L.~Del~Zanna, A.~De~Pace,
  M.~Nardi, G.~Pagliara, V.~Chandra, {A study of vorticity formation in high
  energy nuclear collisions}, Eur. Phys. J. C 75~(9) (2015) 406, [Erratum:
  Eur.Phys.J.C 78, 354 (2018)].
\newblock \href {http://arxiv.org/abs/1501.04468} {\path{arXiv:1501.04468}},
  \href {https://doi.org/10.1140/epjc/s10052-015-3624-1}
  {\path{doi:10.1140/epjc/s10052-015-3624-1}}.

\bibitem{Jiang:2016woz}
Y.~Jiang, Z.-W. Lin, J.~Liao, {Rotating quark-gluon plasma in relativistic
  heavy ion collisions}, Phys. Rev. C 94~(4) (2016) 044910, [Erratum:
  Phys.Rev.C 95, 049904 (2017)].
\newblock \href {http://arxiv.org/abs/1602.06580} {\path{arXiv:1602.06580}},
  \href {https://doi.org/10.1103/PhysRevC.94.044910}
  {\path{doi:10.1103/PhysRevC.94.044910}}.

\bibitem{Deng:2016gyh}
W.-T. Deng, X.-G. Huang, {Vorticity in Heavy-Ion Collisions}, Phys. Rev. C
  93~(6) (2016) 064907.
\newblock \href {http://arxiv.org/abs/1603.06117} {\path{arXiv:1603.06117}},
  \href {https://doi.org/10.1103/PhysRevC.93.064907}
  {\path{doi:10.1103/PhysRevC.93.064907}}.

\bibitem{Vilenkin:1978hb}
A.~Vilenkin, {Parity Violating Currents in Thermal Radiation}, Phys. Lett. B 80
  (1978) 150--152.
\newblock \href {https://doi.org/10.1016/0370-2693(78)90330-1}
  {\path{doi:10.1016/0370-2693(78)90330-1}}.

\bibitem{Vilenkin:1979ui}
A.~Vilenkin, {MACROSCOPIC PARITY VIOLATING EFFECTS: NEUTRINO FLUXES FROM
  ROTATING BLACK HOLES AND IN ROTATING THERMAL RADIATION}, Phys. Rev. D 20
  (1979) 1807--1812.
\newblock \href {https://doi.org/10.1103/PhysRevD.20.1807}
  {\path{doi:10.1103/PhysRevD.20.1807}}.

\bibitem{Vilenkin:1980zv}
A.~Vilenkin, {QUANTUM FIELD THEORY AT FINITE TEMPERATURE IN A ROTATING SYSTEM},
  Phys. Rev. D 21 (1980) 2260--2269.
\newblock \href {https://doi.org/10.1103/PhysRevD.21.2260}
  {\path{doi:10.1103/PhysRevD.21.2260}}.

\bibitem{Son:2009tf}
D.~T. Son, P.~Surowka, {Hydrodynamics with Triangle Anomalies}, Phys. Rev.
  Lett. 103 (2009) 191601.
\newblock \href {http://arxiv.org/abs/0906.5044} {\path{arXiv:0906.5044}},
  \href {https://doi.org/10.1103/PhysRevLett.103.191601}
  {\path{doi:10.1103/PhysRevLett.103.191601}}.

\bibitem{Liu:2017spl}
Y.~Liu, I.~Zahed, {Pion Condensation by Rotation in a Magnetic field}, Phys.
  Rev. Lett. 120~(3) (2018) 032001.
\newblock \href {http://arxiv.org/abs/1711.08354} {\path{arXiv:1711.08354}},
  \href {https://doi.org/10.1103/PhysRevLett.120.032001}
  {\path{doi:10.1103/PhysRevLett.120.032001}}.

\bibitem{Liu:2017zhl}
Y.~Liu, I.~Zahed, {Rotating Dirac fermions in a magnetic field in 1+2 and 1+3
  dimensions}, Phys. Rev. D 98~(1) (2018) 014017.
\newblock \href {http://arxiv.org/abs/1710.02895} {\path{arXiv:1710.02895}},
  \href {https://doi.org/10.1103/PhysRevD.98.014017}
  {\path{doi:10.1103/PhysRevD.98.014017}}.

\bibitem{Chen:2015hfc}
H.-L. Chen, K.~Fukushima, X.-G. Huang, K.~Mameda, {Analogy between rotation and
  density for Dirac fermions in a magnetic field}, Phys. Rev. D 93~(10) (2016)
  104052.
\newblock \href {http://arxiv.org/abs/1512.08974} {\path{arXiv:1512.08974}},
  \href {https://doi.org/10.1103/PhysRevD.93.104052}
  {\path{doi:10.1103/PhysRevD.93.104052}}.

\bibitem{Jiang:2016wvv}
Y.~Jiang, J.~Liao, {Pairing Phase Transitions of Matter under Rotation}, Phys.
  Rev. Lett. 117~(19) (2016) 192302.
\newblock \href {http://arxiv.org/abs/1606.03808} {\path{arXiv:1606.03808}},
  \href {https://doi.org/10.1103/PhysRevLett.117.192302}
  {\path{doi:10.1103/PhysRevLett.117.192302}}.

\bibitem{Ebihara:2016fwa}
S.~Ebihara, K.~Fukushima, K.~Mameda, {Boundary effects and gapped dispersion in
  rotating fermionic matter}, Phys. Lett. B 764 (2017) 94--99.
\newblock \href {http://arxiv.org/abs/1608.00336} {\path{arXiv:1608.00336}},
  \href {https://doi.org/10.1016/j.physletb.2016.11.010}
  {\path{doi:10.1016/j.physletb.2016.11.010}}.

\bibitem{Chernodub:2016kxh}
M.~Chernodub, S.~Gongyo, {Interacting fermions in rotation: chiral symmetry
  restoration, moment of inertia and thermodynamics}, JHEP 01 (2017) 136.
\newblock \href {http://arxiv.org/abs/1611.02598} {\path{arXiv:1611.02598}},
  \href {https://doi.org/10.1007/JHEP01(2017)136}
  {\path{doi:10.1007/JHEP01(2017)136}}.

\bibitem{Chernodub:2017ref}
M.~Chernodub, S.~Gongyo, {Effects of rotation and boundaries on chiral symmetry
  breaking of relativistic fermions}, Phys. Rev. D 95~(9) (2017) 096006.
\newblock \href {http://arxiv.org/abs/1702.08266} {\path{arXiv:1702.08266}},
  \href {https://doi.org/10.1103/PhysRevD.95.096006}
  {\path{doi:10.1103/PhysRevD.95.096006}}.

\bibitem{Wang:2018sur}
X.~Wang, M.~Wei, Z.~Li, M.~Huang, {Quark matter under rotation in the NJL model
  with vector interaction}, Phys. Rev. D 99~(1) (2019) 016018.
\newblock \href {http://arxiv.org/abs/1808.01931} {\path{arXiv:1808.01931}},
  \href {https://doi.org/10.1103/PhysRevD.99.016018}
  {\path{doi:10.1103/PhysRevD.99.016018}}.

\bibitem{Zhang:2018ome}
H.~Zhang, D.~Hou, J.~Liao, {Mesonic Condensation in Isospin Matter under
  Rotation}, Chin. Phys. C 44~(11) (2020) 111001.
\newblock \href {http://arxiv.org/abs/1812.11787} {\path{arXiv:1812.11787}},
  \href {https://doi.org/10.1088/1674-1137/abae4d}
  {\path{doi:10.1088/1674-1137/abae4d}}.

\bibitem{Braguta:2020eis}
V.~V. Braguta, A.~Y. Kotov, D.~D. Kuznedelev, A.~A. Roenko,
  \href{https://doi.org/10.1134/S0021364020130044}{Study of the
  confinement/deconfinement phase transition in rotating lattice su(3)
  gluodynamics}, JETP Letters 112~(1) (2020) 6--12.
\newblock \href {https://doi.org/10.1134/S0021364020130044}
  {\path{doi:10.1134/S0021364020130044}}.
\newline\urlprefix\url{https://doi.org/10.1134/S0021364020130044}

\bibitem{Chen:2020ath}
X.~Chen, L.~Zhang, D.~Li, D.~Hou, M.~Huang, {Gluodynamics and deconfinement
  phase transition under rotation from holography} (10 2020).
\newblock \href {http://arxiv.org/abs/2010.14478} {\path{arXiv:2010.14478}}.

\bibitem{Chernodub:2020qah}
M.~Chernodub, {Inhomogeneous confining-deconfining phases in rotating plasmas}
  (12 2020).
\newblock \href {http://arxiv.org/abs/2012.04924} {\path{arXiv:2012.04924}}.

\bibitem{Yamamoto:2013zwa}
A.~Yamamoto, Y.~Hirono, {Lattice QCD in rotating frames}, Phys. Rev. Lett. 111
  (2013) 081601.
\newblock \href {http://arxiv.org/abs/1303.6292} {\path{arXiv:1303.6292}},
  \href {https://doi.org/10.1103/PhysRevLett.111.081601}
  {\path{doi:10.1103/PhysRevLett.111.081601}}.

\bibitem{Andronic:2017pug}
A.~Andronic, P.~Braun-Munzinger, K.~Redlich, J.~Stachel, {Decoding the phase
  structure of QCD via particle production at high energy}, Nature 561~(7723)
  (2018) 321--330.
\newblock \href {http://arxiv.org/abs/1710.09425} {\path{arXiv:1710.09425}},
  \href {https://doi.org/10.1038/s41586-018-0491-6}
  {\path{doi:10.1038/s41586-018-0491-6}}.

\bibitem{Andronic:2012ut}
A.~Andronic, P.~Braun-Munzinger, J.~Stachel, M.~Winn, {Interacting hadron
  resonance gas meets lattice QCD}, Phys. Lett. B 718 (2012) 80--85.
\newblock \href {http://arxiv.org/abs/1201.0693} {\path{arXiv:1201.0693}},
  \href {https://doi.org/10.1016/j.physletb.2012.10.001}
  {\path{doi:10.1016/j.physletb.2012.10.001}}.

\bibitem{Vovchenko:2014pka}
V.~Vovchenko, D.~V. Anchishkin, M.~I. Gorenstein, {Hadron Resonance Gas
  Equation of State from Lattice QCD}, Phys. Rev. C 91~(2) (2015) 024905.
\newblock \href {http://arxiv.org/abs/1412.5478} {\path{arXiv:1412.5478}},
  \href {https://doi.org/10.1103/PhysRevC.91.024905}
  {\path{doi:10.1103/PhysRevC.91.024905}}.

\bibitem{Hagedorn:1965st}
R.~Hagedorn, {Statistical thermodynamics of strong interactions at
  high-energies}, Nuovo Cim. Suppl. 3 (1965) 147--186.

\bibitem{Cabibbo:1975ig}
N.~Cabibbo, G.~Parisi, {Exponential Hadronic Spectrum and Quark Liberation},
  Phys. Lett. B 59 (1975) 67--69.
\newblock \href {https://doi.org/10.1016/0370-2693(75)90158-6}
  {\path{doi:10.1016/0370-2693(75)90158-6}}.

\bibitem{Andronic:2009gj}
A.~Andronic, et~al., {Hadron Production in Ultra-relativistic Nuclear
  Collisions: Quarkyonic Matter and a Triple Point in the Phase Diagram of
  QCD}, Nucl. Phys. A 837 (2010) 65--86.
\newblock \href {http://arxiv.org/abs/0911.4806} {\path{arXiv:0911.4806}},
  \href {https://doi.org/10.1016/j.nuclphysa.2010.02.005}
  {\path{doi:10.1016/j.nuclphysa.2010.02.005}}.

\bibitem{Ambrus:2014uqa}
V.~E. Ambru\c{s}, E.~Winstanley, {Rotating quantum states}, Phys. Lett. B 734
  (2014) 296--301.
\newblock \href {http://arxiv.org/abs/1401.6388} {\path{arXiv:1401.6388}},
  \href {https://doi.org/10.1016/j.physletb.2014.05.031}
  {\path{doi:10.1016/j.physletb.2014.05.031}}.

\bibitem{Ambrus:2015lfr}
V.~E. Ambrus, E.~Winstanley, {Rotating fermions inside a cylindrical boundary},
  Phys. Rev. D 93~(10) (2016) 104014.
\newblock \href {http://arxiv.org/abs/1512.05239} {\path{arXiv:1512.05239}},
  \href {https://doi.org/10.1103/PhysRevD.93.104014}
  {\path{doi:10.1103/PhysRevD.93.104014}}.

\bibitem{Chen:2017xrj}
H.-L. Chen, K.~Fukushima, X.-G. Huang, K.~Mameda, {Surface Magnetic Catalysis},
  Phys. Rev. D 96~(5) (2017) 054032.
\newblock \href {http://arxiv.org/abs/1707.09130} {\path{arXiv:1707.09130}},
  \href {https://doi.org/10.1103/PhysRevD.96.054032}
  {\path{doi:10.1103/PhysRevD.96.054032}}.

\bibitem{Wheaton:2004qb}
S.~Wheaton, J.~Cleymans, {THERMUS: A Thermal model package for ROOT}, Comput.
  Phys. Commun. 180 (2009) 84--106.
\newblock \href {http://arxiv.org/abs/hep-ph/0407174}
  {\path{arXiv:hep-ph/0407174}}, \href
  {https://doi.org/10.1016/j.cpc.2008.08.001}
  {\path{doi:10.1016/j.cpc.2008.08.001}}.

\bibitem{Cleymans:2005xv}
J.~Cleymans, H.~Oeschler, K.~Redlich, S.~Wheaton, {Comparison of chemical
  freeze-out criteria in heavy-ion collisions}, Phys. Rev. C 73 (2006) 034905.
\newblock \href {http://arxiv.org/abs/hep-ph/0511094}
  {\path{arXiv:hep-ph/0511094}}, \href
  {https://doi.org/10.1103/PhysRevC.73.034905}
  {\path{doi:10.1103/PhysRevC.73.034905}}.

\bibitem{Bazavov:2014pvz}
A.~Bazavov, et~al., {Equation of state in ( 2+1 )-flavor QCD}, Phys. Rev. D 90
  (2014) 094503.
\newblock \href {http://arxiv.org/abs/1407.6387} {\path{arXiv:1407.6387}},
  \href {https://doi.org/10.1103/PhysRevD.90.094503}
  {\path{doi:10.1103/PhysRevD.90.094503}}.

\bibitem{Fukushima:2010is}
K.~Fukushima, {Phase diagram of hot and dense QCD constrained by the
  Statistical Model}, Phys. Lett. B 695 (2011) 387--391.
\newblock \href {http://arxiv.org/abs/1006.2596} {\path{arXiv:1006.2596}},
  \href {https://doi.org/10.1016/j.physletb.2010.11.040}
  {\path{doi:10.1016/j.physletb.2010.11.040}}.

\end{thebibliography}
\bibliographystyle{elsarticle-num}

\end{document}